**An effective electrostatic-confinement based fusion approach**

R. K. Paul*

Department of Applied Physics

Birla Institute of Technology, Mesra

Deoghar Campus

Jasidih-814142, Deoghar, Jharkhand, India.

*Corresponding author

Telephone no. +91-94-30764874

Email: ratan_bit1@rediffmail.com

**Abstract**

The paper reports a new electrostatic-confinement based fusion approach, where, a new non-equilibrium distribution function for an ion-beam, compressed by an external electric force, has been derived. This distribution function allows the system to possess appreciably low and insignificant thermal energy irrespective of the energy per particle. The spread in the energy among the particles is attributed to the collisions in presence of the external force, whereas; for equilibrium, the spreading in energy is due to the absence of the force. The reactivity for a deuterium-deuterium fusion, using the proposed distribution function, has been computed. It is shown that the fusion time is comparable to the energy confinement time, collision time and transit time of the ion for beam energy greater than 160 keV. The estimated energy gain Q (ratio of fusion power to the power consumed by the system) is around 10 for beam energy 160 keV and ion density $10^{18}$ cm$^{-3}$. The energy loss due to particle scattering is estimated and is taken into consideration in the estimation of energy gain. An outline of a possible device is proposed according to the model and is not similar to conventionally used in IEC systems based on collisions of the beam with a reflex beam or with background gas or plasma.

**Keywords:** Classical particles, Compressed beam, Statistical distribution function, Non-equilibrium, Reaction cross-section.

**PACS:** 05.20._y, 05.20.Jj, 05.70.Ce, 05.70.Ln, 25.45._z

## I. INTRODUCTION

Maxwell-Boltzmann distribution function is obtained by maximizing the probability, subject to the condition that the total number of particles N remains fixed and the total energy U associated with the particles remains constant.[1] Departure from this distribution function has also been reported in some cases.[2,3] This departure is attributed to the lack of randomization in collisions among the particles. A procedure to obtain non-equilibrium distribution function using Boltzmann transport equation is to begin with an equilibrium distribution and subject it to a small perturbation.[4] Non-equilibrium distribution function is also obtained as an expansion about a local Maxwellian using Boltzmann transport equation.[5] Computer simulation using non-equilibrium molecular dynamics is also very important to explain nonequilibrium.[6] The prediction of the single particle distribution function using guiding centre theory is also described in a review.[7] In fusion devices, like Tokamak [8], a high temperature (~ 10 keV) Maxwellian plasma at equilibrium is required in order to obtain a necessary two particle scattering cross section. Accordingly, the device becomes complex and bulky. The effort to use non-equilibrium distribution function of the reacting species in an inertial electrostatic confinement of ionized gases has been reported as a relatively simple alternative fusion device.[9]

The aim of the current paper is to formulate and investigate a particular non-equilibrium distribution function and find the conditions, which could result in a physically significant two particles scattering cross section leading to thermonuclear fusion. In the present paper, a differential equation has been formulated by introducing an external force in the total energy equation such that U and N remain constants. Consequently, a new distribution function for a compressed beam of classical particles has been obtained as the solution to the equation.

The novel feature of this distribution function is that, even though the system has a very small thermal energy, the energy per particle can be appreciably large. In section II, a description and the most recent developments of Inertial Electrostatic Confinement is described. In section III the method of formulating aforesaid distribution function and its features are described. The estimation of reactivity is described in Section IV. Time scale of the process is discussed in section V. Particle loss and energy gain are described in section VI and VII respectively. An outline of the device is proposed in section VIII. Section IX concludes the essence of the work.

## II. INERTIAL ELECTROSTATIC CONFINEMENT

Inertial Electrostatic Confinement (IEC) devices usually involve two hollow concentric electrodes where innermost electrode is permeable to charged particle flow.[9-13] By these electrodes arrangement, an electrostatic potential well is established. Ions are accelerated and converged towards the centre by this potential well resulting in a core of higher ion density. If the potential well is sufficiently deep, ions will accelerate to energies sufficient for fusion. The essential features of IEC schemes are (1) electrostatic confinement (2) strong central peaking of the ion density (3) a mono-energetic energy distribution. However, the major problem with IEC schemes is that it requires more energy to maintain their non-thermal plasma distribution than it produces by means of fusion.[14-16] Moreover, power loss in the grid for these devices is in excess with the fusion power.[15] A non-Maxwellian plasma can also be maintained in a more spatially extensive region with two body collisions that thermalize the plasma in one region.[17] A new concept was also suggested to improve electron confinement by means of surrounding an IEC electrostatic potential-well with a polyhedral cusp magnetic field.[18] These electrons form a negative potential well that is capable of ion confinement. Another electrostatic confinement fusion apparatus is reported in which the cross sectional area of plasma confinement zone shrinks to nearly zero dimension at one point, while accelerating the plasma to an energy of 100 keV within that zone.[19] The apparatus has achieved a fusion triple product of $1.5 \times 10^{19} s\, m^{-3} keV$. In another scheme, the focused ion beam used in standard IEC devices is replaced by an oscillating plasma sphere that periodically get compressed to a considerably narrow dimension thereby producing a very high ion density and temperature to induce large scale oscillations inside.[20,21] To avoid the ion bombardment and erosion of the cathode, a reverse polarity IEC device is reported [22] where negative space charge has the potential to produce the electric field required for ion acceleration and confinement without exposing electrodes to the plasma. Recently, spectroscopic studies on IEC devices have also been reported highlighting the ion energy distribution.[23-26] The parameters for IEC reactors utilizing deuteron-based fuels are (1) $\varphi_{well} \cong 300 kV$ (2) $T_i \cong 100\ keV$ (3) $T_e \cong 75\ keV$ (4) $n_i \cong 1 \times 10^{18} cm^{-3}$.[14]

### III. METHODS AND ANALYSIS

Consider the entire phase space consisted of a large number of identical cells. Let the cells be denoted by $a_r$. The cell $a_r$ contains $n_r$ particles each having a definite amount of energy $\varepsilon_r$. The system under consideration is a compressed beam of particles. Its total energy and the number of particles remain unchanged. Due to compression, other cells will be filled up instead of having all particles in a particular cell as in the case of a beam. This will impart energy to the system in the form of heat. On the other hand, energy in the form of work enters the system when external forces due to compression act on the system. Thus the change in internal energy might occur in the following ways; (i) the change in energy due to the work done by the system and (ii) the appearance of heat energy i.e., $dU = -dW + dQ$. Where dU is the change in internal energy, -dW is the work done by the system and dQ is the amount of heat energy produced. Since $dQ = dW$, the change in internal energy dU is zero.

In such a case, the external force should be conservative because the total energy of the system remains constant in the presence of external field change. Hence the beam should be compressed or converged by electric force similar to that in the inertial electrostatic confinement device.[9]

The temperature of the system can be estimated by equating the electrostatic energy $\frac{1}{2}\epsilon_0 E^2$ with the thermal energy $\frac{3}{2}NT$ for a given density N.

$$T = \frac{\varepsilon E^2}{3N} = 1.8 \times 10^{-13} E^2 \text{ eV for } N = 10^{14} \text{ cm}^{-3}$$

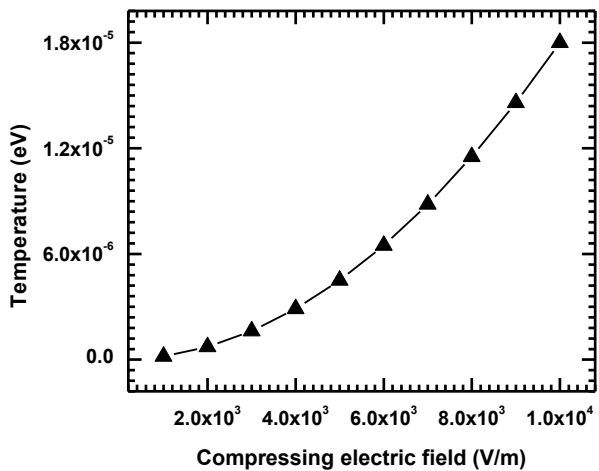

FIG.1. Temperature versus compressing electric field

Fig. 1 is the plot between T and E. The temperature is very low

$[\cong 10^{-7} eV \text{ to } 10^{-1} eV]$ corresponding to the electric field $(\approx 10^3$-$10^6$ Vm$^{-1})$.

The temperature increases as the electric field increases.

As the particle number and the total energy are constants

$$\sum n_r = N \qquad (1)$$

And $\sum n_r \epsilon_r = U = NE_{av}$ \qquad (2)

where, N is the total number of particles, U is the total energy of the system and $E_{av}$ is the beam energy.

again, $\sum \dfrac{n_r}{N} = 1$ \qquad (3)

and $\sum \dfrac{n_r \varepsilon_r}{N} = \dfrac{U}{N}$ \qquad (4)

putting $\omega_r = \dfrac{n_r}{N}$, then

$$\sum \omega_r = 1 \qquad (5)$$

and

$$\sum \omega_r \varepsilon_r = E_{av} \qquad (6)$$

where $E_{av} = \dfrac{U}{N}$

For U and N to be constants, Equation (5) and (6) reduce to

$\sum \delta \omega_r = 0$

This can be written in the limit $\delta \epsilon_r \to 0$ as

$$\sum \frac{d\omega_r}{d\varepsilon_r} \delta\varepsilon_r = 0 \tag{7}$$

And

$$\sum \delta(\omega_r \varepsilon_r) = 0$$

Which can be written in the limit $\delta\epsilon_r \to 0$ (refer to Fig.1 for its justification) as

$$\sum \frac{d}{d\varepsilon_r}(\omega_r \varepsilon_r) \delta\varepsilon_r = 0 \tag{8}$$

In general, the equation (8) can be written as,

$$\sum \left(\omega_r + \varepsilon_r \frac{d\omega_r}{d\varepsilon_r}\right) \delta\varepsilon_r = 0 \tag{9}$$

The second term in equation (9) corresponds to the energy change by changing the particle number due to compression of the beam for which $\varepsilon_r$ is constant. The first term arises because of the energy change by a change in the external coordinates $\lambda$, such that,

$$\omega_r d\varepsilon_r = \frac{n_r}{N} d\varepsilon_r = \frac{n_r}{N} \frac{d\varepsilon_r}{d\lambda} d\lambda,$$

where, $\frac{d\varepsilon_r}{d\lambda}$ represents a force brought to act on the particles of the cell under consideration, $d\varepsilon_r$ being the change in energy of the cell.

From equation (7) and (9),

$$\sum \left\{(1+\beta\varepsilon_r)\frac{d\omega_r}{d\varepsilon_r} + \beta\omega_r\right\} \delta\varepsilon_r = 0 \tag{10}$$

where, $\beta$ is the undetermined multiplier and bears the dimension of inverse of energy.

Since each term in the above summation to be individually zero, we may write

$$\frac{d\omega_r}{d\varepsilon_r} + \frac{\beta\omega_r}{1+\beta\varepsilon_r} = 0 \qquad (11)$$

for a physically valid distribution function where $\frac{d\omega_r}{d\varepsilon_r}$ is always negative. The justification is that, as the second term in relation (11) is evidently positive, $\frac{d\omega_r}{d\varepsilon_r}$ must be negative i.e., either $d\omega_r$ or $d\varepsilon_r$ must be negative with the other one remaining positive. Considering a situation in which all the particles are initially in the energy state $\varepsilon_{r'}$. Due to compression, a number of particles migrate to other cells of energy $\varepsilon_r$. $n_r$ is the number of particles for other cells whereas $n_{r'}$ is the particle number for energy state $\varepsilon_{r'}$. One can write,

$$dU = \sum n_r d\varepsilon_r + n_{r'} d\varepsilon_{r'} + \sum \varepsilon_r dn_r + \varepsilon_{r'} dn_{r'} = 0$$

Since, $dn_{r'}$ is invariably negative and $dn_r$ is always positive therefore, $d\varepsilon_r$ should be negative and $d\varepsilon_{r'}$ should be positive, such that each term in equation (10) is individually zero and $dU$ turns out to be zero as well.

The solution of equation (11) is,

$$\omega_r = \frac{C'}{1+\beta\varepsilon_r} \qquad (12)$$

Hence,

$$n_r = \frac{C'N}{1+\beta\varepsilon_r} \qquad (13)$$

where $C'$ is a normalization constant and $\frac{1}{\beta}$ is the thermal energy.

The number of particles, having energy in between $\epsilon_r$ and $\epsilon_r + d\epsilon_r$ is

$n_r(\epsilon_r)d\epsilon_r = [\rho(\epsilon_r)d\epsilon_r]n_r(\epsilon_r)$, where $\rho(\epsilon_r)d\epsilon_r$ is the number of states that have energies in between $\epsilon_r$ and $\epsilon_r + d\epsilon_r$.

Here the value of $\rho(\epsilon_r)$ is taken constant because the density of states is assumed constant in the small energy interval and zero outside this interval. This is a micro-canonical ensemble and can be very well adapted for representing a system with a fixed energy.[1]

Hence,

$$n_r = \frac{CN}{1+\beta\varepsilon_r}, \text{ where } C = C'\rho(\epsilon_r) \tag{14}$$

The value of C and $\beta$ can be obtained by using the following conditions

$$\int_0^{\varepsilon_c} \frac{CN}{1+\beta\varepsilon_r} d\varepsilon_r = N \tag{15}$$

and

$$\int_0^{\varepsilon_c} \frac{C\varepsilon_r}{1+\beta\varepsilon_r} d\varepsilon_r = \frac{U}{N} = E_{av} \tag{16}$$

where, $\varepsilon_c$ is the upper cut-off energy.

From equation (15) and (16),

$$\log(1+\beta\varepsilon_c) = \frac{\beta}{C} \tag{17}$$

and

$$\frac{C}{\beta^2}\left(\beta\varepsilon_c - \frac{\beta}{C}\right) = E_{av} \tag{18}$$

A guess solution for β and C are,

$$\beta \cong \frac{\exp\left(\frac{\varepsilon_c}{E_{av}}\right) - 1}{\varepsilon_c} \tag{19}$$

and

$$C \cong \frac{E_{av}}{\varepsilon_c}\left(\frac{\exp\left(\frac{\varepsilon_c}{E_{av}}\right) - 1}{\varepsilon_c}\right) \tag{20}$$

With the assumption $\frac{1}{\beta} \ll E_{av}$

From equation (14),

$$n_r = \frac{E_{av}}{\varepsilon_c} N \left[\frac{\exp\left(\frac{\varepsilon_c}{E_{av}}\right) - 1}{\varepsilon_c}\right]\left[\frac{1}{1 + \frac{\left\{\exp\left(\frac{\varepsilon_c}{E_{av}}\right) - 1\right\}\varepsilon_r}{\varepsilon_c}}\right] \tag{21}$$

Relation in equation (21) represents the distribution function for a compressed beam. This non-Maxwellian distribution function, with a sharp cut off energy, occurs despite collisions in the presence of a conservative force. In this case, the force will bring about the random motion among the particles and hence, there is a spreading of energy. This present distribution function is quite different from the non-equilibrium distribution functions as used in [16] and [14], where it has been shown that the power required to maintain these non-equilibrium distribution functions, despite collisions, is larger than the fusion power.

The Maxwellian function can be written as

$$\exp(-\beta\varepsilon_r) = 1/\exp(\beta\varepsilon_r) = 1/(1 + \beta\varepsilon_r + (\beta^2\varepsilon^2_r/2!) + ------$$

When $\beta\varepsilon_r$ is small i.e., $\beta\varepsilon_r \ll 1$, the higher order term can be neglected. This is the condition for which a Maxwellian function reduces to relation (14). On the other hand under the condition, either $\beta \to 0$ (i.e. $1/\beta \to \infty$) or $\varepsilon_r \ll kT$, the Maxwellian function takes to form of relation 14.

The following equation is obtained using equations (18), (19) and (20).

$E_{av} \cong E_{av} - \frac{1}{\beta}$ with the assumption $\frac{1}{\beta} \ll E_{av}$. This implies the condition $\left[exp\left(\frac{\epsilon_c}{E_{av}}\right) - 1\right] \gg \epsilon c E a v$.

After substituting $\frac{1}{\beta}$ from equation (19)

$$E_{av} \cong \left[\frac{E_{av}}{\{exp\left(\frac{\epsilon_c}{E_{av}}\right)-1\}}\right]\left[\left\{exp\left(\frac{\epsilon_c}{E_{av}}\right) - 1\right\} - \frac{\epsilon_c}{E_{av}}\right] \tag{22}$$

The solution of the above equation, i.e. $\epsilon_c = 0$ is however, not physically acceptable due to the following reasoning. For $\epsilon_c = 0$, $\frac{1}{\beta}$ is undefined; but for $\epsilon_c \to 0$, $\frac{1}{\beta} = E_{av}$, which does not satisfy the equation $E_{av} \approx E_{av} - \frac{1}{\beta}$. No definite value of $\frac{\epsilon_c}{E_{av}}$ can be obtained using the condition $\left[exp\left(\frac{\epsilon_c}{E_{av}}\right) - 1\right] \gg \frac{\epsilon_c}{E_{av}}$, which implies $\epsilon_c \gg E_{av}$. However, an approximate value can be estimated using equation (19), where $\frac{1}{\beta} \cong 10^{-6}$ (refer to fig.1) and for $E_{av} = 2000 eV$, $\epsilon_c \cong 52429\ eV$ yielding $\frac{\epsilon_c}{E_{av}} \cong 26$.

The method, in principle, is based on the variation in energy in contrast to the usual Maxwell Boltzmann method, where the variation in particle number within each cell is considered. The present method is also used to obtain Maxwell Distribution function for the case of a system where a small thermodynamical process, such as compression, is carried out so slowly that it is reversible. The compression is necessary to provide the variation in energy of the system. As in the previous case the Equations (7) & (9) are also applicable in this case i.e.

$$\sum \frac{d\omega_r}{d\varepsilon_r}\delta\varepsilon_r = 0 \qquad (23)$$

and

$$\sum \left(\omega_r + \varepsilon_r \frac{d\omega_r}{d\varepsilon_r}\right) \delta\varepsilon_r = 0 \qquad (24)$$

The first term corresponds to the change in the energy due to the work done by the system. The force will have no impact on the second term in eq. (24), since, the changes in $n_r$ are governed by the random motions of the system and are not affected by changes in the external coordinates. Hence eq. (24) can be written as

$$\sum \omega_r \delta\varepsilon_r = 0 \qquad (25)$$

From eq.(23) and eq.(25)

$$\sum \left(\frac{d\omega_r}{d\varepsilon_r} + \beta\omega_r\right) \delta\varepsilon_r = 0 \qquad (26)$$

Therefore,

$$\frac{d\omega_r}{d\varepsilon_r} + \beta \omega_r = 0 \tag{27}$$

The solution of eq.(27) is Maxwellian

$$\omega_r = \frac{n_r}{N} = A e^{-\beta \varepsilon_r} \tag{28}$$

Due to compression by the external conservative force the system does work and the work in the form of energy enters in to the system. This energy is compensated by the occurrence of random energy as a result of the distribution of total beam energy U among the particles due to collision. Thereby the total energy of the system remains constant (vide equation (16)). The mean energy of the distribution is $E_{av} = \frac{U}{N}$ and the thermal energy is $\frac{1}{\beta}$ whereas in case of Maxwell-Boltzmann distribution mean energy equals to thermal energy $\frac{1}{\beta}$.

From equation (19)

$$\frac{\frac{1}{\beta}}{E_{av}} = \frac{Thermal\ energy\ of\ each\ particle}{Directed\ energy\ of\ each\ particle} = 7.45 \times 10^{-11}$$

Fig.2 shows the variation of $1/\beta$ with $E_{av}$ (in eV) when $\varepsilon_c$ (in eV) is obtained using the relation (22).

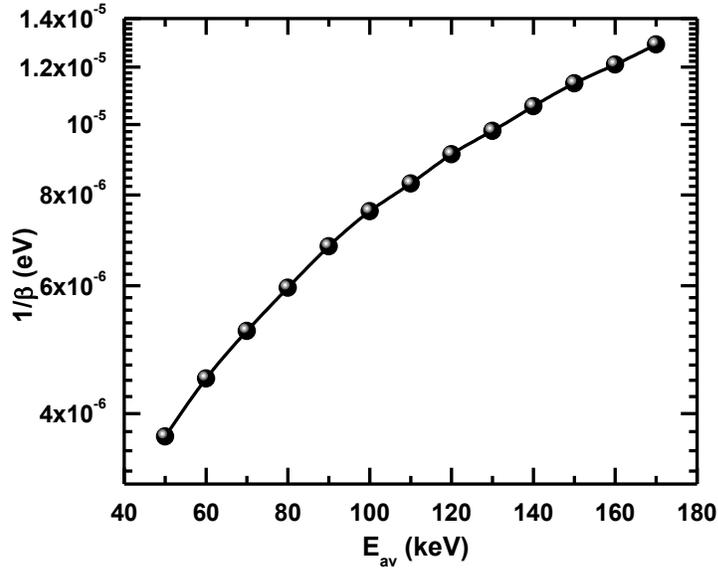

**FIG.2.** Variation of $1/\beta$ with $E_{av}$.

It has been observed that the value of $1/\beta$ increases with the increase in $E_{av}$. This indicates that the system possesses a very small ($\approx 10^{-6} eV$) thermal energy, whereas energy per particle can be very high. This spreading of energy is due to collisions in the presence of force and is found to be similar as shown in Fig. 1. On the other hand, the temperature for the standard IEC device is very high ($T = \frac{2}{3}E = 100\ keV$) in comparision to the present case ($\cong 10^{-6} eV$). However, this cannot be explained on the basis of energy uncertainty and an estimate of average $d\varepsilon_r$ is also obtained by using the following two equations,

$$\iiint dxdydz \iiint dp_x dp_y dp_z = 2\pi V(2m)^{\frac{3}{2}} \varepsilon_r^{\frac{1}{2}} d\varepsilon_r = h^3 \tag{29}$$

where, h is the Planck's constant and V is the volume occupied by the particles

$$\text{and } \overline{d\varepsilon_r} = h^3 \left[ \frac{1}{2\pi V(2m)^{\frac{3}{2}}} \right] \left( \frac{\int_0^{\varepsilon_c} \frac{n_r d\varepsilon_r}{\varepsilon_r^{\frac{1}{2}}}}{\int_0^{\varepsilon_c} n_r d\varepsilon_r} \right) \tag{30}$$

where, $n_r$ is given by equation (21).

Comparing the value of $\frac{1}{\beta}$ obtained by this model with $\overline{d\varepsilon_r}$ for a compressed proton beam in which the values of V and $E_{av}$ are taken as: $V = 1.5 \times 10^{-6}$ m$^3$ and $E_{av}$ = 3000 eV, the value of $\overline{d\varepsilon_r}$ is found to be around $1.8 \times 10^{-29}$ eV, which is not in good agreement with $\frac{1}{\beta}$.

From equation (20)

$$C = 5.04 \times 10^8 \times \frac{1}{E_{av}}$$

Fig. 3 shows the variation of C with $E_{av}$. The value of C decreases with the increase in $E_{av}$. The upper cut off energy $\varepsilon_c$ increases with the increase in $E_{av}$. Hence, the value of C decreases with the increase in $E_{av}$.

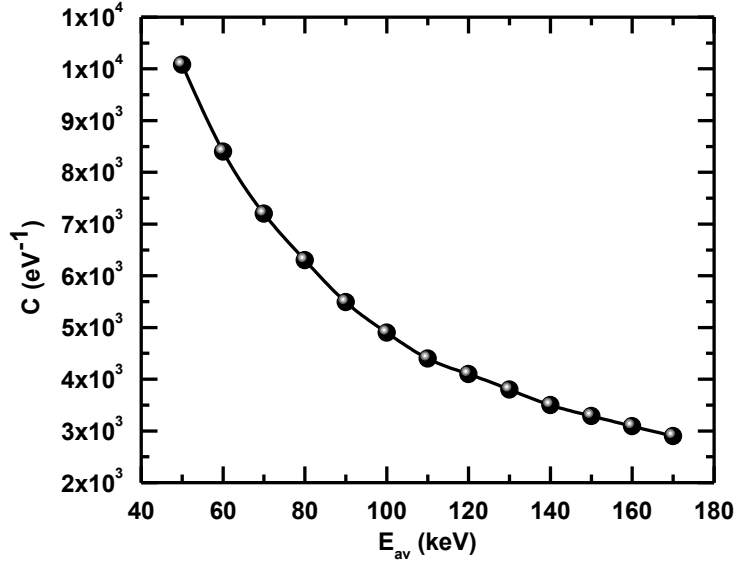

**FIG. 3.** Variation of C with $E_{av}$.

## IV. REACTIVITY:

The estimation of reactivity $<\sigma v>$, for an equilibrium Maxwellian system and a beam, is available in the literature[27,28,29]. Fusion reactivity is also estimated for a spherically convergent ion focus[30]. The reactivity is computed by using the following equation.

$$<\sigma v> = \iint f(v_1)f(v_2)\sigma(v)v\, d^3v_1 d^3v_2 \qquad (31)$$

Where f(v) is the distribution function, $\sigma(v)$ is the reaction cross section and $v$ is the relative velocity.

Equation (1) is written in the following form using the derived distribution function.

$$<\sigma v>$$
$$= \frac{1}{2}c^2 m^2 \int_0^{v_c}\int_0^{v_c}\int_0^{\pi} \frac{v_1}{1+\frac{mv_1^2}{2T}(v_1^2+v^2+2vv_1\cos\theta)^{\frac{1}{2}}} \cdot \frac{1}{1+\frac{m}{2T}(v_1^2+v^2+2vv_1\cos\theta)} \cdot \sigma(v)v^3 \sin\theta\, d\theta\, dv\, dv_1$$

$$(32)$$

Here $g(v)dv = \frac{Cmvdv}{1+\frac{m}{2T}v^2}$, $T = \frac{1}{\beta}$ and $v_c = (\frac{2\epsilon_c}{m})^{\frac{1}{2}}$

The value of $\sigma$ as a function $v$ for deuterium ion is taken from the relation[27].

$$\sigma(v) = \frac{1.79 \times 10^{-23}}{v^2} \exp\left(-\frac{14.578}{v}\right) \tag{33}$$

$$\sigma(v) = 10^{-27} \exp\left(-9.082 + 8.9606v - 2.59541v^2 + 0.386918v^3 - 0.0287342v^4 + 0.0084047v^5\right) \tag{34}$$

where, $v$ is the relative velocity measured in units of $10^8$ cm sec$^{-1}$ and equation (33) is valid below 80 keV and equation (34) is valid for energies between 80 keV and 1000 keV.

The value of $<\sigma v>$ is estimated numerically and its variation with beam energy $E_{av}$ is shown in Fig. (4). Its value is small for beam energy up to 130 keV and increases for higher energy beam. This can be attributed by the fact that the number of ions in the tail of the distribution is large. Hence, the probability of reaction per unit volume and per unit time is large. The typical value of reactivity for $E_{av}$=160 keV is $3.39 \times 10^{-12} cm^3 s^{-1}$.

.

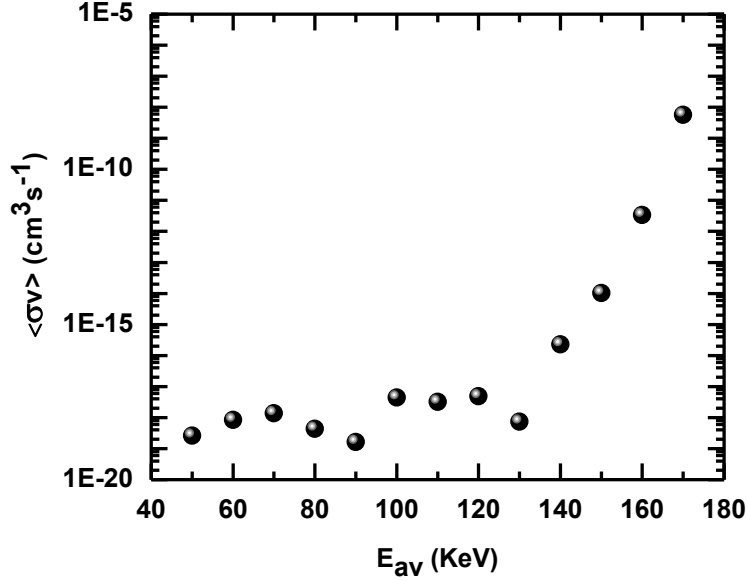

**FIG.4.** Variation of reactivity with the beam energy $E_{av}$.

The reaction cross section for IEC reactor utilizing deuteron based fuels is $0.495 \times 10^{-16} cm^3 s^{-1}$.[14]

## V. TIME SCALES:

### A. Energy confinement time

The confinement time is defined [8] as $\tau_E = \frac{\frac{3}{2}NT}{P}$ where P is the total input power. The input power is written as $P = iV$ where $i$ is the beam current and $V$ is the applied voltage. The current for a cylindrical beam of radius r is given as $i = Nqv\pi r^2$ where $v$ is the velocity of the ion, $q$ is the ionic charge per particle and r is the radius of the beam. Hence, $P = iV = Nqv\pi r^2 V = \pi r^2 NvE_{av}$ where $E_{av} = qV$ is the beam-energy. The velocity $v$ can be expressed as $v = \frac{\sqrt{m}}{\sqrt{2}\pi r^2 \sqrt{E_{av}}}$. Therefore, $\tau_E = \frac{\sqrt{m}}{\sqrt{2E_{av}}} \times \frac{\pi r^2 l}{\pi r^2} = \frac{l\sqrt{m}}{\sqrt{2}\sqrt{E_{av}}}$ where $T = \frac{2}{3}E_{av}$ is considered and $l$ is taken as beam length. It is observed that the value of $\tau_E$ decreases as the beam energy increases. For $E_{av} = 160 keV$ and l=1 m, the value of $\tau_E$ is found to be $3.0 \times 10^{-7}s$.

**B: Fusion Time**: It is the characteristic time of interaction by a test ion with another ion and is defined as $\tau_F = \frac{1}{N<\sigma v>}$. For $E_{av}$=160 keV and N=$10^{18}$ cm$^{-3}$, the typical value of fusion time is $2.94 \times 10^{-7}s$.

**C: Transit Time:** It is the time an ion takes to cross the physical length in between the electrodes. The transit time is estimated using $\tau_t = \frac{l}{v}$, where $l$ is the beam length and $v$ is the velocity of the ion. For $E_{av}$=160 keV and $l$=1m the transit time is $3.5 \times 10^{-7}s$.

**D: Collision Time:** The ion- ion collision time is estimated by using the following relation $eE = \frac{eV}{l} = \frac{mv}{\tau_i}$, where $\tau_i$ is the collision time and $v$ is the velocity of the ion. For $E_{av}$=160 keV, and $l$=1m the typical value of collision time is $3.28 \times 10^{-7}s$.

The collision time is comparable to the transit time. Hence, thermalization of the beam particles occurs within the transit time. Also, the energy confinement time is nearly equal to the transit time. Fig.(5) shows the variation of fusion time and energy confinement time with the beam energy. Fusion time is greater than the energy confinement time when beam energy is less than 150 keV. For $E_{av}$=160 keV and N=$10^{18}$ cm$^{-3}$, $\frac{\tau_E}{\tau_F} \cong 1$. This indicates that Deuterium-Deuterium fusion process is only possible when the beam energy is greater than 160 keV under the proposed scheme.

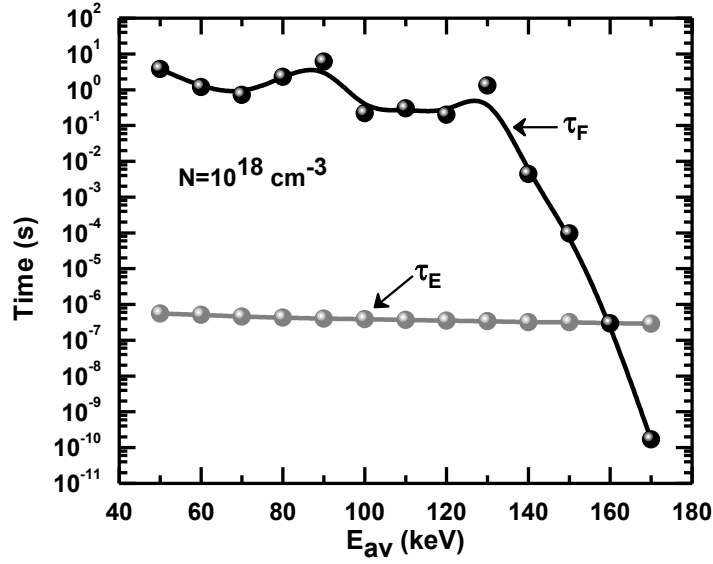

**Fig. 5.** Variation of confinement time and fusion time as a function of beam energy.

## VI. Particle loss

A fraction of particle is scattered due to collision above the cutoff energy $\varepsilon_c$ and lost from the system. For a steady state this particle loss is continuously replenished by ion souece. Particle loss from electrostatic wells has been studied in the references [31, 32]. Expression for particle loss rate from electrostatic well in the small mirror ratio approximation (pR<1) is [32]

$$\dot{n} = -\frac{4N\exp{(-p)}}{\tau\pi^{1/2}} \times [1 - pR \int_0^1 dt\, (1-t)^{\frac{1}{2}}\exp{(-pRt)}]$$

Where for pR→1 the integral in the large parentheses equals 0.46. And

$p = \frac{Ze\varphi_O}{T}$, where $ze\varphi_0 = E_{av}$ and $T = \frac{2}{3}E_{av}$ is considered.

Therefore, $p = \frac{3E_{av}}{2E_{av}} = 1.5$.

The collision time $\tau$ is defined as [32]

$$\tau \cong \frac{M^{\frac{1}{2}} T^{\frac{3}{2}}}{2^{\frac{1}{2}} \pi e^4 z^4 N \ln\Delta}$$

For N=$10^{18}$ cm$^{-3}$ and $E_{av} = 160\ keV$, $\dot{n} \cong -5.26 \times 10^{21}$ in case of a deuterium atom. Therefore, within energy confinement time $\tau_E$ the particle loss from the system is estimated as $N_L = 1.57 \times 10^{15}$. Hence the approximate energy loss ($\dot{n}\tau_E \epsilon_c$) is $1.57 \times 10^{15} \times 26.6 \times 16 \times 10^4 \times 1.6 \times 10^{-19} = 1.069 \times 10^3 J cm^{-3}$. This estimation shows that power loss due to particle scattering is significant and should be taken into consideration in the energy gain calculation. For $E_{av}$=170 keV and N=$10^{16}$ cm$^{-3}$, particle loss rate within the energy confinement time is $1.39 \times 10^{11}$. Hence the energy loss is 0.1 $j\ cm^{-3}$.

## VII. ESTIMATION OF ENERGY GAIN

The energy gain is the ratio of the energy produced per unit volume to the input energy per unit volume of the system. The energy produced per unit volume is given by $W_P = \tau_E P_f$ where $P_f = \frac{N_s^2}{2} <\sigma v> E_f$. For $E_{av} = 160$ keV, $N_s = N - N_L = (10^{18} - 1.57 \times 10^{15}) = 0.99 \times 10^{18} cm^{-3}$, $<\sigma v> = 3.39 \times 10^{-12} cm^3 s^{-1}$ ( as taken from Fig.4) and $E_f = 3.2 \times 10^6$ eV( Power released per fusion), $W_P = 2.6 \times 10^5 j cm^{-3}$. The input energy is the beam energy $W_r = N E_{av} = 25.6 \times 10^3 j cm^{-3}$. Again energy loss due to particle loss is $W_{loss} = 1.069 \times 10^3 j cm^{-3}$. Hence, the energy gain $Q = \frac{W_P}{W_r + W_{loss}} \cong 10$.

For $E_{av}$=170 keV and N=$10^{16}$ cm$^{-3}$, The fusion time $\tau_F \cong 1.75 \times 10^{-8} s$ for $<\sigma v> \cong 5.69 \times 10^{-9} cm^3 s^{-1}$ when $\tau_E \cong 2.9 \times 10^{-7} s$. Hence $\frac{\tau_F}{\tau_E} \cong 6 \times 10^{-2}$. The value of $W_P$ is $42 \times 10^3 j cm^{-3}$ and $W_r = 272\ j cm^{-3}$. Again energy loss due to particle loss is $W_{loss} = 0.1\ j\ cm^{-3}$. Hence the energy gain is $Q \cong 150$.

The energy gain Q of the device is appreciable for beam energy beyond 160 kev. Operating regimes with $Q > 100$ for penning fusion system have also been identified[33].

## VIII. PROPOSED DEVICE:

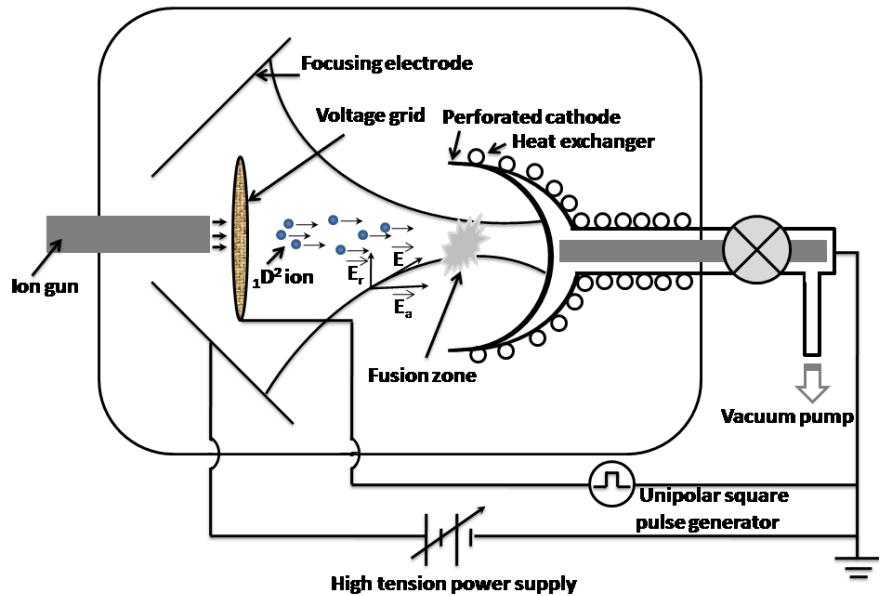

FIG. 6. A schematic of the proposed device.

The apparatus, maintained at ultra high vacuum, is conceptually similar to one used for the generation of carbon nanotubes using a focusing electrostatic electric field[34,35] and is schematically shown in Fig. 6. It is, in practical, not similar to the conventional IEC devices, which essentially have relied upon collisions of the beam with a reflex beam[9,10] or with background gas[36,37] or plasma[33,19].

In the proposed apparatus, deuterium ion beam will be introduced from an ion gun and will subsequently be subject to an initial acceleration by using a voltage grid connected to a high-voltage DC square pulse generator. A truncated cone shaped focusing electrode, connected to a high-voltage DC source will then guide and compress the beam as it proceeds towards the grounded cathode. In the schematic $\vec{E_a}$ is the accelerating component of the electrostatic focusing field and $\vec{E_r}$ is the component responsible for compressing the beam. By suitably tuning the voltages (of the grid and the focusing electrode) and the orientation of the focusing electrode, one can ideally create a situation, where fusion will take place at the focal plane of the hemispherical cathode in the vicinity of the beam axis according to the proposed model. During operation, the cathode will be grounded and subsequently cooled below the temperature corresponding to the work-function of the cathode material. This prevents mixing of the beam with any unwanted electron. After thermalization, the scattered particles will immediately be taken away from the reactor with the help of a suitable vacuum system as depicted in Fig.6.

## IX. CONCLUSIONS:

A new statistical distribution function for a cylindrical beam of classical particles compressed by applying an external electric field has been derived. It is found that the spreading of the energy among the particles is due to the collisions subject to the condition that the beam is under the influence of a compressing force. The distribution function, thus generated, shows that even though the system has a very small thermal energy (~$10^{-6}$ eV), the number of particles towards the high energy side becomes appreciably large. The D-D fusion reactivity estimated using this function is reasonably higher for the beam energy greater than 160 keV. On the other hand, the fusion time is also comparable to the energy confinement time and transit time above 160 keV beam energy. The reactivity that quantifies the fusion yield is predicted to be higher above this beam energy provided the compressed beam is subjected to follow the distribution function described in the present paper. The particle loss due to scattering is computed and is taken into consideration for the estimation of energy gain. The estimated energy gain of the device working on the principle described in the present article is predicted to be appreciable (Q=10) for the ion density greater than $10^{18}$ cm$^{-3}$. An outline of a possible device, based on the principle described in the present paper, is proposed.

**Acknowledgements:** The author is grateful to the administration of BIT, Mesra for its support and encouragement in accomplishment of this work. The author also gratefully acknowledges the help and support of R. Kumar (BIT, Muscat), S. Gupta (Central University of Jharkhand, Ranchi), M. K. Sinha (BIT, Mesra), K Bose (BIT, Mesra) and Soumen Karmakar (BIT, Deoghar).